\documentclass[conference]{IEEEtran}
\IEEEoverridecommandlockouts
\usepackage{cite}
\usepackage{amsmath,amssymb,amsfonts}
\usepackage{algorithmic}
\usepackage[pdftex]{graphicx}
\usepackage{textcomp}
\usepackage{xcolor}
\usepackage[]{hyperref}
\usepackage{enumitem}
\usepackage{longtable}
\usepackage{float}
\usepackage{makecell}
\usepackage{tcolorbox}
\usepackage{pifont}
\usepackage{url}
\usepackage{orcidlink}
\usepackage{array}
\usepackage{booktabs}
\usepackage[table,dvipsnames]{xcolor}
\usepackage[utf8]{inputenc}
\usepackage[font=footnotesize,skip=5pt]{caption}
\usepackage{marvosym} % \Letter command

\def\BibTeX{{\rm B\kern-.05em{\sc i\kern-.025em b}\kern-.08em
    T\kern-.1667em\lower.7ex\hbox{E}\kern-.125emX}}

\hypersetup{
    colorlinks=true,
    linkcolor=blue,     
    citecolor=blue,      
    filecolor=blue,     
    urlcolor=blue     
}

\newcommand{\projectsite}{\url{https://github.com/CorporatePhishingStudy}}
\newcommand{\ccross}{\ding{55}}
\newcommand{\ccheck}{\ding{52}}

\begin{document}

\title{Sustaining Cyber Awareness: The Long-Term Impact of Continuous Phishing Training and Emotional Triggers}
% \title{Sustaining Cyber Awareness: \\
% \large The Long-Term Impact of Continuous Phishing Training and Emotional Triggers}

% ----- EDIT AUTHORS/AFFILIATIONS AS NEEDED -----
\author{
\IEEEauthorblockN{Rebeka T\'{o}th \orcidlink{0009-0000-9574-1896}}
\IEEEauthorblockA{\textit{University of Oslo}\\
Oslo, Norway\\
\url{rebekat@uio.no}}
\and
\IEEEauthorblockN{Richard A.\ Dubniczky \orcidlink{0009-0003-3951-1932}\textsuperscript{(\Letter)}}\thanks{\textsuperscript{\Letter} Corresponding author.}
\IEEEauthorblockA{\textit{E\"otv\"os Lor\'and University}\\
Budapest, Hungary} \url{richard@dubniczky.com}
\and

\IEEEauthorblockN{Olga Limonova \orcidlink{0009-0001-7385-7166}}
\IEEEauthorblockA{\textit{University of Luxembourg}\\
Luxembourg, Luxembourg\\
\url{olga.limonova.001@student.uni.lu}}
\and
\IEEEauthorblockN{Norbert Tihanyi \orcidlink{0000-0002-9002-5935}}
\IEEEauthorblockA{\textit{Technology Innovation Institute}\\
 Abu Dhabi, UAE\\
\url{norbert.tihanyi@tii.ae}}
}

\maketitle
\begin{abstract}
Phishing constitutes more than 90\% of successful cyberattacks globally, remaining one of the most persistent threats to organizational security. Despite organizations tripling their cybersecurity budgets between 2015 and 2025, the human factor continues to pose a critical vulnerability. This study presents a 12-month longitudinal investigation examining how continuous cybersecurity training and emotional cues affect employee susceptibility to phishing. The experiment involved 20 organizations and over 1,300 employees who collectively received more than 13,000 simulated phishing emails engineered with diverse emotional, contextual, and structural characteristics. Behavioral responses were analyzed using non-parametric correlation and regression models to assess the influence of psychological manipulation, message personalization, and perceived email source. Results demonstrate that sustained phishing simulations and targeted training programs lead to a significant reduction in employee susceptibility, halving successful compromise rates within six months. Additionally, employee turnover introduces measurable fluctuations in awareness levels, underscoring the necessity of maintaining continuous training initiatives. These findings provide one of the few long-term perspectives on phishing awareness efficacy, highlighting the strategic importance of ongoing behavioral interventions in strengthening organizational cyber resilience. In order to support open science, we published our email templates, source code, and other materials at \projectsite

\end{abstract}

\smallskip


\section{Introduction}
\label{sec:introduction}
Phishing is a type of cyberattack where criminals trick people into revealing sensitive information, such as passwords or financial details, by pretending to be a trustworthy source~\cite{phishing}. Phishing remains one of the most dominant attack vectors in the modern cybersecurity landscape.
According to the United States \textit{Cybersecurity and Infrastructure Security Agency} (CISA), more than 90\% of successful cyberattacks begin with phishing~\cite{CISA2023}. Unlike purely technical exploits, phishing attacks primarily exploit the \textit{human element}, which continues to be the most vulnerable component of organizational security systems~\cite{human_factor1,human_factor2}. While organizations continue making significant investments into strengthening technical safeguards such as firewalls, intrusion detection systems, and AI-driven anomaly detection, adversaries increasingly bypass these mechanisms by manipulating employees directly. This is often achieved through carefully crafted messages that exploit emotions, cognitive biases, and contextual cues to provoke unsafe actions such as clicking on malicious links, opening infected attachments, or disclosing sensitive information~\cite{DODGE200773}. Understanding the behavioral mechanisms underlying phishing susceptibility therefore constitutes a central challenge for cybersecurity research \cite{Abroshan2021}.

In 2022, Lain et al.~\cite{phishinglargescale2022} conducted one of the largest and most comprehensive phishing studies to date, involving over 14,000 employees over a 15-month period. The simulated phishing campaign was carried out from July 2019 to October 2020. One of the most surprising findings was that voluntary embedded (contextual) phishing training\footnote{In voluntary training, users who fall for a phishing email are directed to a training page; however, participation and completion are neither mandatory nor monitored}, a widely adopted industry practice, proved ineffective and, in some cases, even increased employees’ susceptibility to phishing attacks. The authors noted that their study did not evaluate mandatory training and suggested that future research should explore whether enforced interaction with training materials could yield better outcomes. 

Addressing this gap, the present study examines the effectiveness of \textit{mandatory embedded phishing training} and investigates its impact on employee phishing resilience. In our scenario, when a user falls for a phishing email, an automated training session is generated and assigned to them in the HR system as mandatory training.

This study analyzes employee responses to phishing in a real corporate environment over 12 months, drawing on data from more than 1,300 participants and incorporating psychological and contextual factors such as personalization, message origin, and emotional cues like urgency and authority. The study also assesses how repeated exposure to phishing simulations influences long-term resilience, evaluating the long-term training effect in real-world security awareness programs. 

\subsection{Main contribution}

By combining longitudinal data with a strong theoretical foundation, this research identifies which manipulative tactics are most effective and which training approaches best enhance lasting user awareness. This work thereby contributes to resolving open questions previously noted in the literature~\cite{DODGE200773,phishinglargescale2022}.

Our study is guided by the following research questions, designed to explore how different psychological, contextual, and organizational factors influence phishing susceptibility and to assess the impact of mandatory embedded training on long-term user resilience.
\begin{tcolorbox}[colback=gray!10,   colframe=black,   arc=6pt,     boxrule=0.7pt,  left=2mm, right=2mm, top=1mm, bottom=1mm,
    rounded corners]
\begin{itemize}
    \item \textbf{RQ1:} How does continuous phishing simulation and \textit{mandatory} training influence the frequency of unsafe employee actions?
    \item \textbf{RQ2:} How effective are repeated phishing tests, immediate feedback, and training cycles at reducing subsequent successful attacks?
    \item \textbf{RQ3:} How do emotional and contextual cues influence the likelihood of a successful phishing attack?
\end{itemize}
\end{tcolorbox}
To address these questions, this work makes the following key contributions:

\begin{itemize}
\item Empirical evidence from an extensive email-based training campaign comprising over 13,000 simulated phishing attempts conducted over a 12-month period across 20 organizations.
\item A collection of phishing email templates, each carefully annotated with relevant emotional and contextual cues.
\item A comprehensive analytical framework, including the complete codebase for data parsing, preprocessing, and statistical analysis of phishing susceptibility patterns.
\end{itemize}

To support open science and reproducibility, the dataset of email templates and the  scripts for statistical analysis are publicly available at \projectsite.

\subsection{Ethical Considerations}
The phishing campaign was conducted within a European holding company (hereafter referred to as the \textit{"holding company"}). This study formed part of the holding's emerging security awareness program and fully complied with applicable data protection regulations, including the GDPR. Data collection was performed under the organization’s legitimate interest in enhancing information security and employee awareness. All analyses were carried out exclusively on anonymized and aggregated data, ensuring that no personally identifiable information is accessible to the public. Employees were notified that simulated phishing exercises form an integral component of the corporate training framework. The study protocol underwent internal review by the firm's data protection officer, who confirmed its compliance with all relevant ethical and legal standards.

\section{Background and Related Work}
In this section, we review related work on emotional cues and phishing tactics to understand the key factors that make phishing emails so effective.

Recent studies have demonstrated that phishing effectiveness is strongly linked to the psychological mechanisms it exploits. Attackers often rely on emotional triggers, such as fear, urgency, curiosity, altruism, or authority, to manipulate recipients into bypassing rational decision-making and acting impulsively \cite{Jayatilaka2024,Musuva2019,Wang2023}. For example, messages framed around immediate financial loss or urgent corporate compliance have been shown to be particularly persuasive. Khadka et al.~\cite{Khadka2024} further revealed that the emotional tone and framing context of an email significantly affect the likelihood of unsafe user behavior. De Bona and Paci observed that classic phishing cues like urgency and authority appear less effective, likely because users have become habituated to these tactics \cite{DeBona2025}. This raises questions about how phishing tactics evolve and adapt to user awareness over time.

Alongside the identification of psychological manipulation strategies, research has emphasized the importance of training and awareness in reducing phishing susceptibility. Traditional lecture-based cybersecurity education often struggles to produce lasting behavioral change. In contrast, interactive approaches such as gamified training, role-playing exercises, and embedded reminders within daily work tasks have been shown to be significantly more effective \cite{Chen2024,Lain2024b,Rahartomo2025}. These experiential methods allow employees to simulate real-world phishing encounters, improving recognition skills and developing practical coping mechanisms. Beyond immediate awareness gains, such interventions have the potential to enhance long-term resilience against evolving social engineering tactics.

Despite notable progress, several challenges continue to constrain phishing-related research. Kavvadias and Kotsilieris highlight that many empirical investigations into phishing resilience exhibit methodological shortcomings, such as short study durations, artificial laboratory settings, and limited or non-representative participant samples \cite{Kavvadias2025}. These factors hinder the external validity and generalizability of findings to real-world organizational contexts.

Moreover, the role of demographic variables—including age, gender, and technological proficiency—remains ambiguous across existing literature. While some studies identify these characteristics as potential predictors of phishing susceptibility, other works find only marginal or insignificant effects \cite{Heijden2019,Waqas2023}. Lain et al. reported correlations between age, computer skills, and phishing susceptibility, but found no such relationship for gender, which contradicts earlier findings \cite{contradict}.

An important question remains unanswered: to what extent are untrained organizations vulnerable at baseline, and how effectively can targeted  \textit{mandatory} training initiatives strengthen their resilience over the span of a one-year intervention? The following sections outlines our methodology and explores the answers to our research questions.
\begin{figure*}[t] 
\centering
\includegraphics[width=\textwidth]{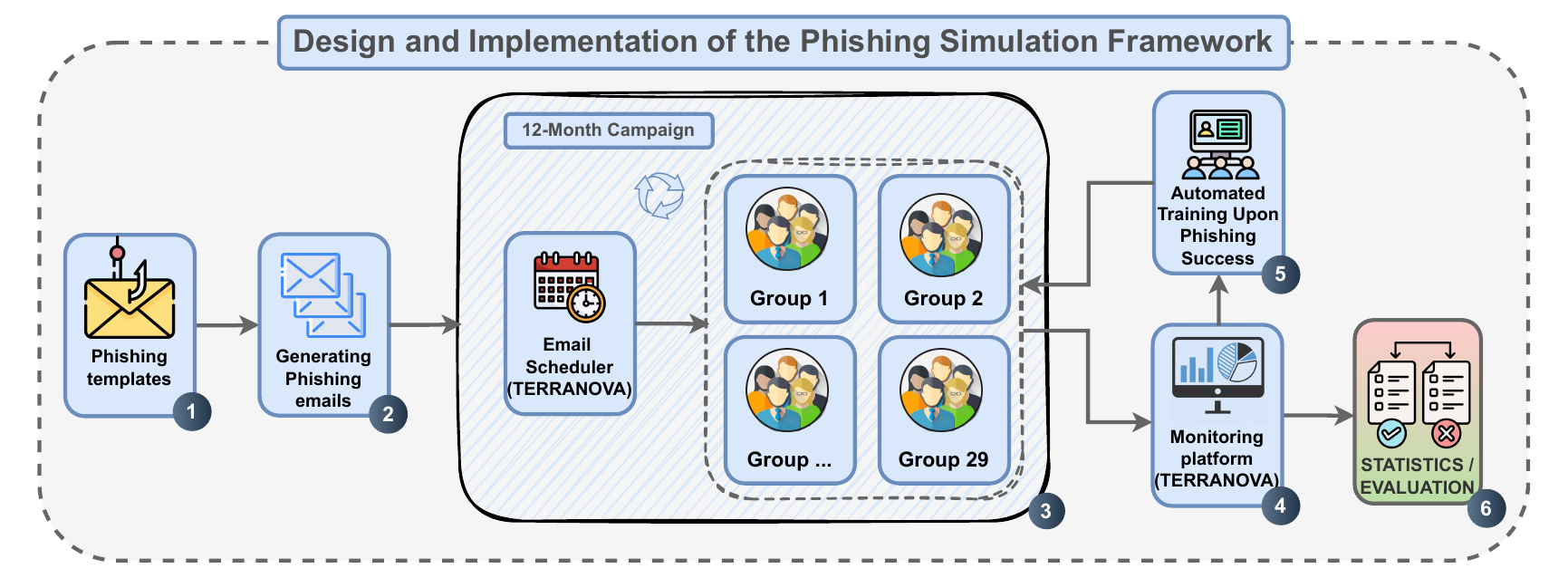}
\caption{Overview of the full research workflow, illustrating each stage from phishing email template design through the email campaign, data collection, trainings, and statistical analysis.}
\label{fig:framework}
\end{figure*}
\section{Methodology}
\label{sec:methodology}

In this section we will detail our methodology while conducting the study and drawing the conclusions. Figure ~\ref{fig:framework} provides a structured overview of the methodological steps, summarizing the experimental design.

\subsection{Organizational Context}

The study was conducted in collaboration with the 20 European subsidiaries of the holding company. The holding encompasses multiple subsidiaries operating under distinct brands across a broad range of online and offline sectors, including digital media, subscription services, e-commerce, payment and billing solutions, advertising technology, and audiovisual production. 
These subsidiaries maintain a high degree of operational independence and preserve their distinct brand identities. However, they rely on shared corporate infrastructure, including email systems, identity management, and security awareness platforms. This structure provides an ideal context for analyzing phishing susceptibility, given the prevalence of internal communication and cross-brand correspondence. To preserve confidentiality and adhere to internal disclosure policies, the holding company has elected to remain anonymous in this publication.

It is important to note that the organization enforces a mandatory vacation period in July and August. During this time, most employees are on leave, leading to a significant reduction in email activity. Consequently, phishing exposure and response rates during these months may be lower than in typical operational periods. Additionally, new employees usually join the holding company in two main cohorts, one in May and another in late summer or early autumn. This onboarding schedule causes predictable changes in the number of staff members who have not yet received phishing test emails.

\subsection{Sample Group}

The study involved a little over 1,300 participants, 73\% of whom were men and 27\% women, aged between 18 and 58 years. The participants represented diverse professional roles across departments, ensuring that the sample was representative of typical corporate staff who are potential phishing targets, including engineering, management, finance, HR, legal, and other roles.

Participants were distributed across multiple geographical regions in Central and Western Europe and North America, with a smaller number of employees working remotely from other international locations. To preserve the validity of the data, members of the IT Security Department as well as others who were aware of the content of the upcoming tests were excluded, as their insight might skew the results.

During the study period, the total participant number fluctuated slightly, as some employees joined or left the companies. Nonetheless, all employees who were active at any point in 2024 were included, resulting in a total of more than 1{,}300 participants. This design ensured ecological validity by capturing the dynamics of a real corporate environment rather than a static laboratory sample \cite{Waqas2023}.

\subsection{Cybersecurity Awareness and Training Program}

Throughout the study period, the holding operated a comprehensive cybersecurity awareness program mandatory for all employees. Each new hire was required to complete an introductory e-learning module, followed by a short assessment consisting of ten multiple-choice questions. A minimum score of eight correct responses was necessary to pass and gain access to internal systems; this certification was renewed annually from then on through a repeat of the same training. In addition to the onboarding and annual online courses, the holding company conducted two organization-wide awareness sessions during the year, held in June and October. These sessions addressed current security threats, corporate policies, and best practices in safe online conduct, including phishing awareness.

The program further incorporated an adaptive learning component focused on phishing resilience. Employees who engaged unsafely with simulated phishing messages automatically received a prompt to complete a short corrective module emphasizing the identification of common phishing indicators and illustrating real-world attack examples. Completion of this remedial mini-course was \textit{mandatory before resuming regular work activities}. It is important to note that this constitutes a significant aspect of our methodology, as Lain et al.~\cite{phishinglargescale2022} demonstrate that voluntary (non-mandatory) training has no measurable effect in their studies. This multilayered and responsive training framework provided continuous reinforcement of secure practices and delivered immediate, contextual feedback following unsafe interactions.

\subsection{Psychological Impact of Cyberattacks Model}

This research builds on the \textit{Psychological Impact of Cyberattacks Model} proposed by Bogdanov, Rahartomo, and Smith, which provides a structured framework for analyzing how psychological and contextual elements embedded in phishing messages influence user behavior \cite{Bogdanov2018,Rahartomo2025b,Smith2025}. The model integrates emotional manipulation strategies, contextual framing, and user characteristics, thereby offering a comprehensive perspective on phishing susceptibility. In our study, we operationalize specific elements of this model that are both empirically measurable and relevant to the available dataset. These elements fall into three primary categories:

\begin{enumerate}
    \item \textbf{Email Presentation and Contextual Framing}
    \begin{itemize}
        \item \textit{Personalization:} Whether the email explicitly includes the recipient's name or is addressed in a generic manner.
        \item \textit{Attack Source:} Whether the email appears to originate from an internal corporate system or from an external organization.
    \end{itemize}
    \item \textbf{Emotional Manipulation}
    \begin{itemize}
        \item \textit{Emotions:} Targeted emotional states, such as fear, curiosity, altruism, or greed, that the phishing message attempts to evoke.
        \item \textit{Intensifiers:} Amplifying cues such as urgency (e.g., “immediate action required”) or authority (e.g., instructions from a senior executive).
    \end{itemize}
    \item \textbf{User Characteristics}
    \begin{itemize}
        \item \textit{Demographics:} In this study, due to the anonymous nature of the collection methods, we did not measure results based on user characteristics.
    \end{itemize}
\end{enumerate}

While the original model includes additional variables such as cognitive processing style and prior exposure, these could not be systematically measured in the current context. Our focus is therefore on the most practical and observable dimensions within a real corporate environment, namely email presentation, contextual framing, and emotional manipulation.

\subsection{Email Templates}

Prior to the experiment, we created a total of 31 core phishing email templates. From these templates, thousands of distinct emails can be generated at any time. Each template was made in English and reflected common manipulation tactics observed in real-world phishing campaigns. The design process was guided by the \textit{Psychological Impact of Cyberattacks Model} psychological model \cite{Bogdanov2018,Rahartomo2025b}, ensuring that templates systematically varied across key psychological factors:

\begin{itemize}
    \item \textbf{Emotional triggers:} fear, urgency, curiosity, altruism, and greed.
    \item \textbf{Personalization:} contains the user's name, phone number, information related to their manager, or any other personal information.
    \item \textbf{Attack source:} sender appears to be within the organization or not (via a spoofed domain).
    \item \textbf{Intensifiers:} urgency and authority were also incorporated into selected emails to test their amplifying effects.
\end{itemize}

The final set of phishing emails is detailed in Table~\ref{tab:phishing_emails}, along with their effectiveness during the study. For distribution and behavioral tracking, we used the \textit{Terranova} security awareness platform \cite{Abroshan2021}. This platform enabled a controlled simulation of phishing attacks and precise recording of user interaction, including link clicks, file downloads, and form data submissions.

\begin{table*}[t]
\centering
\caption{Characteristics of phishing emails used in the simulated attacks ordered by their success rate.}
\label{tab:phishing_emails}

\rowcolors{3}{gray!20}{white} % 
\begin{tabular}{|c|p{5.5cm}|ccccl|cc|}

\toprule
\multicolumn{2}{c}{} & \multicolumn{5}{c}{\textbf{Contextual Cues}} & \multicolumn{2}{c}{\textbf{Statistics}} \\
\cmidrule(lr){3-7} \cmidrule(lr){8-9} 
\textbf{Rank} & \textbf{Email Template} & \textbf{Personalized} & \textbf{Urgent} & \textbf{Authoritive} & \textbf{Internal} & \textbf{Emotions} & \textbf{\makecell{Targeted \\ Groups}} & \textbf{\makecell{Success \\ Rate}} \\

\hline

1 & [JIRA] *NAME* mentioned you on RM-48526 & \ccheck & \ccross & \ccheck & \ccross  & F, C, A & 2 & 36.0\% \cellcolor{red!72} \\
2 & DocumentSignature – New Document & \ccross & \ccheck & \ccheck & \ccheck  & C, A & 1 & 12.5\% \cellcolor{red!25} \\
3 & HR Personal Information Update & \ccheck & \ccross & \ccheck & \ccheck  & C, A & 23 & 10.8\% \cellcolor{red!22} \\
4 & RE: Quick approval / Call follow-up & \ccheck & \ccheck & \ccross & \ccheck  & C, F, A & 20 & 10.6\% \cellcolor{red!21} \\
5 & Place your order to receive free company merch! & \ccross & \ccheck & \ccheck & \ccheck  & C, G & 8 & 10.3\% \cellcolor{red!20} \\
6 & You have been added to our Careers Site! & \ccross & \ccross & \ccheck & \ccross  & C, G & 24 & 7.3\%  \cellcolor{red!15} \\
7 & Signature requested on Policy review & \ccross & \ccross & \ccheck & \ccross  & C & 14 & 7.2\% \cellcolor{red!14} \\
8 & Holiday Calendar & \ccheck & \ccross & \ccheck & \ccheck  & F, G & 11 & 6.5\% \cellcolor{red!13} \\
9 & HR News – Save on your taxes  $\dots$ & \ccross & \ccross & \ccheck & \ccheck  & C, G & 20 & 6.3\% \cellcolor{red!12} \\
10 & Password reset / Change password & \ccross & \ccheck & \ccross & \ccross  & F & 18 & 6.2\% \cellcolor{red!12} \\
11 & Halloween Raffle for all employees! & \ccross & \ccheck & \ccheck & \ccheck  & G & 9 & 5.5\% \cellcolor{red!11} \\
12 & Important news! Company collective insurance & \ccross & \ccross & \ccheck & \ccheck  & C, G & 21 & 5.5\% \cellcolor{red!11} \\
13 & Job Invitation from Google & \ccheck & \ccross & \ccheck & \ccross  & C, G & 4 & 4.6\% \cellcolor{red!9} \\
14 & VPN Service Changed & \ccross & \ccheck & \ccross & \ccross  & C, F & 19 & 3.6\% \cellcolor{red!7} \\
15 & Unusual sign-in activity & \ccross & \ccheck & \ccheck & \ccross  & F & 12 & 2.9\% \cellcolor{red!6} \\
16 & Action Required – Unused Vacation Days $\dots$ & \ccheck & \ccheck & \ccheck & \ccheck  & F, G& 9 & 2.9\% \cellcolor{red!6} \\
18 & Instagram copyright infringement notice $\dots$ & \ccheck & \ccheck & \ccheck & \ccross  & F  & 4 & 2.7\% \cellcolor{red!3}\\
19 & Policy Breach Detected  & \ccheck & \ccheck & \ccross & \ccross  & F & 14 & 2.7\% \cellcolor{red!3} \\
17 & You have 2 new messages in Teams & \ccheck & \ccheck & \ccheck & \ccross  & C, A & 18 & 2.6\% \cellcolor{red!5} \\
21 & Happy Holidays! Enjoy this Gift! $\dots$ & \ccross & \ccross & \ccheck & \ccheck  & G & 10 & 2.3\% \cellcolor{red!5} \\
20 & Adobe Enterprise Sync & \ccross & \ccross & \ccross & \ccheck  & C & 3 & 2.2\% \cellcolor{red!4} \\

23 & Google Drive – File pending approval & \ccross & \ccross & \ccheck & \ccross  & C & 24 & 1.8\% \cellcolor{red!4} \\
22 & Black Friday Deals & \ccross & \ccross & \ccross & \ccheck  & C, G & 10 & 1.6\% \cellcolor{red!3} \\
24 & Activate your FREE LinkedIn Learning account & \ccross & \ccheck & \ccheck & \ccross  & C, G & 3 & 0.9\% \cellcolor{red!2} \\
25 & Can you assist me with this? & \ccheck & \ccross & \ccross & \ccheck  & C, A & 13 & 0.8\% \cellcolor{red!2} \\
26 & Microsoft Outlook Encrypted Message & \ccross & \ccross & \ccheck & \ccross  & C & 19 & 0.6\% \cellcolor{red!1} \\
27 & Undeliverable – Email Delivery Blocked & \ccross & \ccross & \ccheck & \ccross  & C, F & 4 & 0.6\% \cellcolor{red!1} \\
28 & Zoom Webinar & \ccross & \ccross & \ccheck & \ccross  & C & 4 & 0.6\% \cellcolor{red!1} \\
29 & Register now – CorporateGPT & \ccross & \ccross & \ccheck & \ccross  & C, G & 5 & 0.5\% \cellcolor{red!1} \\
30 & DHL Shipment Notification & \ccross & \ccross & \ccheck & \ccross  & C, G & 1 & 0.0\% \cellcolor{red!0} \\
31 & Alert – Your Info Was Found on the Dark Web & \ccross & \ccheck & \ccross & \ccross  & C, F & 2 & 0.0\% \cellcolor{red!0} \\

\bottomrule
\end{tabular}
\smallskip

\footnotesize
Legend: \textbf{F} = Fear; \textbf{C} = Curiosity; \textbf{A} = Altruism; \textbf{G} = Greed;
\end{table*}

\subsection{Procedure}

Participants were randomly assigned to 30 groups to ensure balanced group sizes while reducing the likelihood of intra-departmental communication that could compromise the simulations. This design minimized the risk of employees warning one another about ongoing phishing tests. During regular training sessions, participants were instructed to report any suspicious messages directly to the IT Security Department, in accordance with standard corporate security policies.

The phishing simulation was conducted over a 12-month period, from January to December 2024. The initial round of phishing emails was distributed in January, before any mandatory training activities, to establish a reliable baseline. Each participant received one simulated phishing email per month, sent exclusively during standard working hours on business days to maximize engagement. In total, more than 13,000 phishing emails were distributed during the study. Table~\ref{tab:phishing_emails} summarizes the email templates and corresponding failure rates. No participant was exposed to the same phishing template more than once, and each participant was only targeted a maximum of 10 times over the 12-month period to reduce predictability. The emails were also sent at a randomized time during the month, sometimes outside regular business hours.

When a participant engaged in unsafe behavior—such as clicking a malicious link, downloading an attachment, or submitting personal information—they immediately received a follow-up awareness email. This message identified the simulation, explained the indicators that made the email suspicious, and directed the participant to complete a mandatory training. This “just-in-time” intervention provided immediate, context-specific feedback and reinforced learning, an approach demonstrated to decrease the likelihood of repeated unsafe actions \cite{Chen2024,Lain2024b,Rahartomo2025b}.

\subsection{Data Analysis}

Behavioral responses and manipulation variables were systematically analyzed using the \textit{IBM SPSS Statistics 23} software suite, complemented by custom Python scripts for preprocessing and visualization. The analysis followed a structured three-stage workflow to ensure the reliability, validity, and reproducibility of the findings across the large dataset.

\begin{figure*}[htb]
    \centering
    \includegraphics[width=\textwidth]{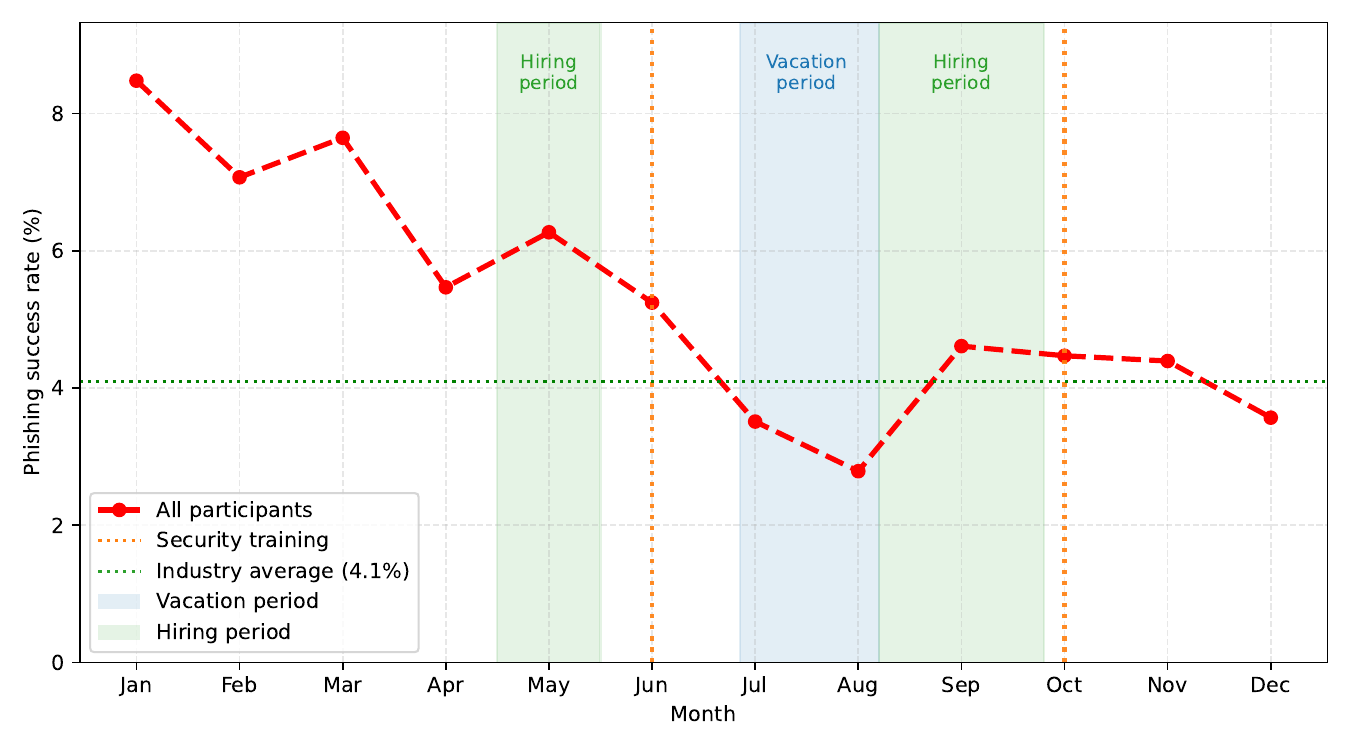}
    \caption{Percentage of successful phishing attempts each month over 12 months for all 20 companies. The industry average is measuring companies that have ongoing training programs (KnowBe4 report \cite{knowbe4_2025_phishing}).}
    \label{fig:success_over_time}
\end{figure*}

\begin{enumerate}
    \item \textbf{Normality testing:} Kolmogorov–Smirnov tests were applied to examine whether the data followed a normal distribution. Since behavioral data in phishing research often exhibit non-normal characteristics, this step was instrumental in determining the suitability of non-parametric methods.
    \item \textbf{Correlation analysis:} Spearman’s rank correlation was employed to investigate associations between psychological manipulation factors (e.g., emotions, intensifiers, source, and personalization) and unsafe user behaviors.
    \item \textbf{Multiple comparison correction:} Given the number of psychological factors tested simultaneously, Bonferroni corrections were applied to adjust significance thresholds.
\end{enumerate}

We used descriptive statistics and cross-tabulations to summarize unsafe actions across conditions.

Behavioral outcomes in phishing simulations are discrete (e.g., clicked, opened attachment, or ignored) and typically non-normally distributed. Therefore, we opted to use non-parametric statistics, as they are robust to deviations from normality and to unequal group sizes. We used Spearman’s rank correlation for associations, with Bonferroni corrections for multiple comparisons.

To verify robustness, we used additional mixed-effects logistic regression models estimated with random intercepts for participant and month, predicting unsafe behavior from emotional and contextual factors. These models yielded results consistent with the non-parametric analyses, confirming that the main conclusions were not dependent on the chosen method.

We developed custom Python scripts to standardize, clean, and analyze the dataset before statistical evaluation. Data were transposed into normalized tables suitable for non-parametric and regression analyses using libraries such as \texttt{pandas} and \texttt{numpy}. Statistical summaries and exploratory visualizations were generated with \texttt{matplotlib}, enabling rapid inspection of behavioral trends across phishing templates and user attributes. Additionally, Python was used to automate the generation of correlation matrices, plots, and regression diagnostics, ensuring full reproducibility of the analytical workflow. All scripts were tested for consistency and are available alongside the dataset on the project’s public repository to support transparent scientific reproducibility.

\subsection{Sample Size and Power}

The full employee population ($N > 1{,}3oo$) participated, yielding over 13{,}000 phishing email exposures. Post-hoc sensitivity analysis indicated sufficient statistical power ($>0.9$) to detect small effects ($|\rho| \approx 0.03$) at $\alpha = 0.05$.

Uncontrolled factors such as departmental norms, tenure, or local communication culture may have influenced results. Randomized group allocation and month-level modeling reduced—but did not fully eliminate—these potential confounds.

\section{Discussion and Results}
\label{sec:results}

In this section, we present the findings of our quantitative analysis and discuss their implications. Overall, the results indicate a clear reduction in phishing susceptibility throughout the 12-month period, with the final rate approaching the industry benchmark of 4.1\% \cite{knowbe4_2025_phishing}.

\subsection{Training Effect on Phishing Susceptibility}

The annual progression of phishing success rates is depicted in Figure~\ref{fig:success_over_time}. Initial results from January show a relatively high compromise rate of 8.5\%, which decreased steadily to a minimum of 2.8\% in August. During the final quarter, the rate stabilized at an average of 4.2\%, only 0.1 percentage points (pp) above the industry average for organizations implementing continuous training initiatives.

As anticipated, notable fluctuations correspond with the organization’s hiring and vacation periods. During onboarding phases, when new employees constitute less than 10\% of the workforce, they account for approximately 25\% of all successful phishing interactions. Conversely, during the summer vacation months, average susceptibility fell to 3.1\%, roughly 1 pp below the industry benchmark. This reduction aligns with reduced overall activity levels, as approximately 25\% of employees were on leave in any given week, with near-complete absence during one week in August. A considerable portion of emails delivered during this time were either ignored or interacted with only after employees’ return. 

The observed downward trend in unsafe behavior over the year demonstrates a robust and sustained training effect. Regular exposure combined with immediate feedback significantly enhanced phishing awareness and reduced impulsive responses. These findings underscore the effectiveness of continuous, simulation-based awareness programs supported by just-in-time corrective feedback.

\begin{figure}[H]
    \centering
    \includegraphics[width=\linewidth]{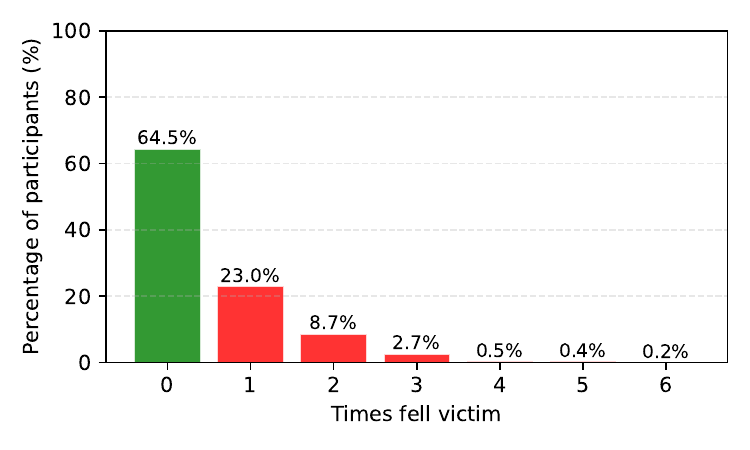}
    \caption{Distribution of participants by the number of unsafe actions recorded during phishing simulations. No individual engaged in unsafe behavior more than six times.}
    \label{fig:victim_count_distribution}
\end{figure}

Our analyses further reveal that the interactive training framework—comprised of simulated emails followed by immediate feedback—produced a 52\% reduction in phishing susceptibility within a six- to eight-month time span. As shown in Figure~\ref{fig:victim_count_distribution}, approximately 70\% of participants who fell for a phishing attempt once did not repeat the unsafe behavior. This improving trend persisted across subsequent incidents, with similar proportional declines between the second and third occurrences. Overall, 64.5\% of employees never engaged in unsafe actions, while 23\% did so only once. The maximum number of repeated unsafe actions observed was six, recorded for just 0.2\% of participants. At least 35.5\% of employees fell for a phishing attempt at least once, which is consistent with the findings of Lain et al.~\cite{phishinglargescale2022}.

\subsection{Effect of Contextual and Emotional Cues on Phishing Success}

\begin{figure}[!htbp]
    \centering
    \includegraphics[width=\linewidth]{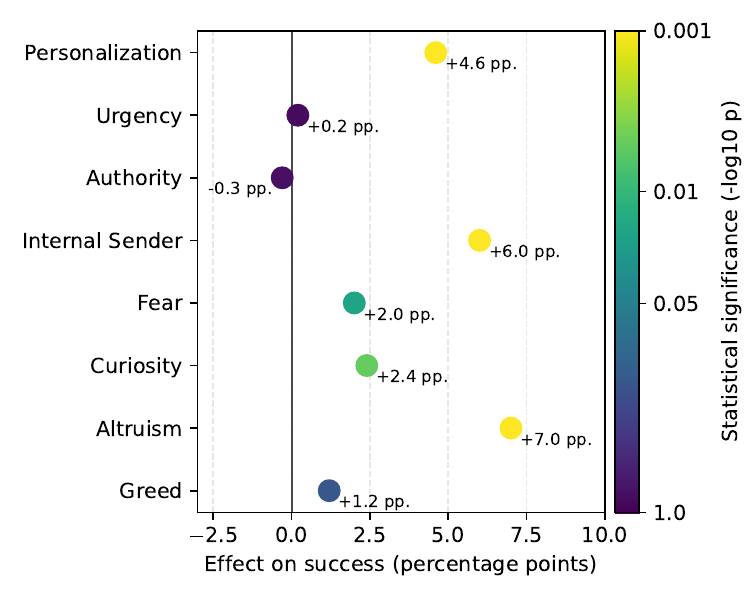}
    \caption{Effect of contextual and emotional cues on phishing success compared to baseline calculated using Spearman's rank correlation.}
    \label{fig:emotional_cue_effects}
\end{figure}

To assess the influence of emotional and contextual cues on user susceptibility, we analyzed the correlation between phishing success rates and eight distinct message-level characteristics using Spearman’s rank correlation. The results are visualized in Figure~\ref{fig:emotional_cue_effects}. 

The analysis included over 13{,}000 simulated phishing events conducted throughout 2024, of which 5.1\% resulted in unsafe user actions. Given the non-normal distribution of behavioral data, Spearman’s rho served as an appropriate non-parametric metric to capture the monotonic relationships between cue presence and user response.

The results reveal that specific emotional cues modestly but significantly affected phishing success. Messages containing \texttt{altruism} appeal demonstrated the strongest positive correlation with unsafe behavior (\(\rho = 0.070, p < 10^{-15}\)), indicating that altruistic or prosocial framing was particularly effective at eliciting user engagement. This aligns with prior research suggesting that appeals to empathy or helpfulness often bypass users' critical scrutiny. Similarly, messages categorized as \texttt{internal source} (\(\rho = 0.060, p < 10^{-11}\)) — those appearing to originate from within the organization — were also associated with elevated phishing success. These findings highlight the increased credibility of intra-organizational messages and demonstrate that both emotional resonance and perceived source authenticity play substantial roles in influencing user trust.

Other cues exhibited smaller yet statistically significant effects. Phishing emails with \texttt{personalized} content (\(\rho = 0.046, p < 10^{-7}\)) correlated with higher compromise rates, suggesting that tailored content remains an effective manipulation strategy even in highly trained environments. Likewise, \texttt{curiosity} and \texttt{fear} cues showed weaker positive associations (\(\rho = 0.024, p = 0.005\); \(\rho = 0.020, p = 0.018\), respectively), implying that emotionally charged or attention-grabbing messages continue to exert influence, though less consistently than altruistic cues. By contrast, \texttt{Greed} — appealing to financial gain — framing did not significantly predict success (\(p = 0.155\)), and messages invoking \texttt{authority} or \texttt{urgency} exhibited no meaningful relationship to unsafe behavior (\(p > 0.7\)). The negligible or negative correlations for these latter cues suggest that heightened user familiarity with urgent or authority-based messages may have reduced their manipulative effectiveness over time.

Taken together, these results demonstrate that while emotional intensity generally influences phishing susceptibility, its effectiveness varies substantially by cue type. Altruistic and internal framing strategies were particularly impactful, confirming that social trust and cooperative intent remain exploitable human factors. Conversely, traditional fear- or urgency-based appeals appear to be losing persuasive power, possibly reflecting growing organizational resilience and awareness through repeated training. This nuanced pattern underscores the importance of tailoring phishing education to emphasize not only overt “red flags,” such as urgency, but also subtle manipulative techniques relying on trust, helpfulness, and perceived legitimacy.

While individual emotional and contextual cues demonstrated statistically significant relationships with phishing success, the effect sizes were relatively modest, with all correlation coefficients below 0.1. This indicates that no single cue, in isolation, exerts a dominant influence on user susceptibility. Instead, phishing effectiveness appears to emerge from the cumulative interaction of multiple subtle manipulations rather than the strength of any single emotional trigger. All the contextual information and emotional triggers sustained a similar rate of descent in effectiveness over the 12 months, indicating that the trainings are effective against all the cues we tested.

To better understand this interplay, we examined combined cue configurations—specifically, scenarios where multiple persuasive elements co-occur within the same message. The analysis shows that emails incorporating two or more high-impact cues, such as altruism combined with an internal email source and personalization framing, achieved markedly higher compromise rates. In the most effective combination of cues, success rates reached approximately a 15\% increase in the relative likelihood of compromise compared to emails that contained none of the identified emotional or contextual indicators.

These findings show that while individual cues exhibit limited predictive power when considered independently, their synergistic use significantly amplifies phishing effectiveness. The results suggest that successful phishing attempts often rely on exploiting multiple complementary psychological mechanisms—such as trust, empathy, and familiarity—within a single message.

\section{Limitations and Threats to Validity}

Although this study provides valuable insights into long-term phishing susceptibility, it has several limitations. First, the research was conducted within a diversified European holding company. While this structure offers a realistic corporate environment with varied roles and contexts, it may not fully capture the diversity of security cultures, technologies, or training practices present across different industries and organizational sizes, for instance, ones in the Americas or Asia.

Second, the study’s reliance on simulated phishing campaigns intrinsically limits the generalizability of results to real-world attacks. Participants may have been more cautious knowing that training exercises were introduced to their work environment, potentially leading to lower overall susceptibility. Furthermore, the content and design of simulated emails, though realistic, cannot perfectly replicate the sophistication or novelty of targeted spear-phishing attempts encountered outside experimental conditions.

Finally, the dataset reflects dynamic workforce changes, such as employee turnover and onboarding cycles, which introduce variability not entirely controllable through the research design. Although these conditions contribute to ecological validity, they may also obscure causal relationships between training exposure and behavioral outcomes.

\section{Conclusion}

In this longitudinal study, we ran a 12-month-long phishing experiment on more than 1,300 employees of 20 companies using 31 email templates. We provide a detailed examination of employee susceptibility to phishing within a large European corporate environment. By combining continuous simulation-based training with detailed behavioral analysis, we demonstrate that consistent exposure and immediate feedback can significantly reduce unsafe actions. Phishing success rates were nearly halved within the first six months and stabilized close to industry benchmarks, underscoring the long-term effectiveness of sustained awareness initiatives.  

Our findings highlight that while emotional and contextual cues influence phishing susceptibility, their effects are modest when considered individually. The most impactful attacks combine multiple persuasive elements—particularly signals of internal origin, personalization, and altruistic framing—resulting in success rates nearly 15\% higher than the least effective configurations. These results suggest that contemporary phishing threats increasingly rely on subtle psychological interplay rather than overt urgency or fear tactics.

From a practical standpoint, this research reinforces the importance of adaptive and continuous awareness programs that mirror real-world threats. Organizations should focus on embedding just-in-time training and scenario-based exercises that address compound manipulative techniques rather than isolated patterns.

Our findings further demonstrate that immediate feedback following phishing failures serves as a powerful reinforcement mechanism. Employees who received such feedback were, on average, 70\% less likely to repeat unsafe actions in subsequent simulations, highlighting the efficacy of real-time learning interventions in promoting lasting behavioral change.

Based on these findings, we answer our research questions the following way:

\begin{itemize}
    \item \textbf{RQ1:} How does continuous phishing simulation and mandatory training influence the frequency of unsafe employee actions? \\
    \textbf{Answer:} \textit{The rate of unsafe employee actions declined by roughly half within the first six months of the training campaign, decreasing from 8.5\% to 4.2\%. Thereafter, it remained stable and closely aligned with the industry average observed in organizations implementing comparable awareness programs (4.1\%). This represents a substantial reduction, particularly among employees who initially exhibited higher susceptibility to phishing attempts.}
    \item \textbf{RQ2:} How effective are repeated phishing tests, immediate feedback, and training cycles at reducing subsequent successful attacks? \\
    \textbf{Answer:} \textit{Repeated phishing simulations combined with immediate feedback and targeted remedial training proved highly effective in reducing subsequent successful attacks. Employees who failed a phishing test and completed the follow-up training were, on average, 70\% less likely to repeat unsafe actions in later simulations. This result demonstrates the strong reinforcement effect of just-in-time feedback, significantly enhancing retention and long-term behavioral change among previously susceptible individuals.}
    \item \textbf{RQ3:} How do emotional and contextual cues influence the likelihood of a successful phishing attack? \\
    \textbf{Answer:} \textit{Emotional and contextual cues were found to influence phishing success, though their individual effects were relatively modest. Messages framed around altruistic appeals, internal communication, or personalized context showed the strongest correlations with unsafe actions, increasing success rates by up to 15\% when combined. In contrast, fear- and urgency-based cues exhibited minimal impact, suggesting that subtle trust- and empathy-based manipulations are currently more effective than overt pressure tactics in eliciting user engagement.
}
    
\end{itemize}

\section{Further Research}

Future research should explore how much training is sufficient before reaching diminishing returns, whether the initial improvements in employee vigilance rebound or plateau over time, and how long it takes for learned behaviors to decline without reinforcement. Addressing these questions will be instrumental for developing evidence-based frameworks that balance training intensity with lasting user awareness.

Another major step involves systematically evaluating the effectiveness of periodic organization-wide training sessions. Although our observations did not reveal significant short-term changes following such events, the limited number of test instances prevents conclusive interpretation. Controlled longitudinal experiments comparing different training frequencies—for example, quarterly versus biannual programs—could clarify whether periodic reinforcement enhances, sustains, or merely overlaps with the benefits of individualized feedback-based learning.

Additionally, future research should explore the interaction between personalized training intensity and message complexity. Combining adaptive learning algorithms with behavioral analytics may allow tailoring interventions to employee risk profiles. Expanding this line of inquiry across various industries and cultural contexts could further validate the generalizability of our findings.

\section*{Acknowledgement}

This research is supported and funded by ZEISS Digital Innovation and the Technology Innovation Institute (TII), Abu Dhabi. Additional support is provided by the TKP2021-NVA Funding Scheme under Project TKP2021-NVA-29; ELTE-OTP Cyberlab—a collaboration between Eötvös Loránd University (ELTE) and OTP Bank Plc. None of the organizations mentioned were involved in the phishing simulation campaign.

\bibliographystyle{IEEEtran}
\bibliography{main}

\end{document}